# Maximum Likelihood Estimation for Finite Mixtures of Canonical Fundamental Skew $t$-Distributions: the Unification of the Unrestricted and Restricted Skew $t$-Mixture Models

Sharon X. Lee, Geoffrey J. McLachlan
Department of mathematics, the University of Queensland,
Brisbane, Australia

**Abstract**

In this paper, we present an algorithm for the fitting of a location-scale variant of the canonical fundamental skew $t$ (CFUST) distribution, a superclass of the restricted and unrestricted skew $t$-distributions. In recent years, a few versions of the multivariate skew $t$ (MST) model have been put forward, together with various EM-type algorithms for parameter estimation. These formulations adopted either a restricted or unrestricted characterization for their MST densities.

In this paper, we examine a natural generalization of these developments, employing the CFUST distribution as the parametric family for the component distributions, and point out that the restricted and unrestricted characterizations can be unified under this general formulation. We show that an exact implementation of the EM algorithm can be achieved for the CFUST distribution and mixtures of this distribution, and present some new analytical results for a conditional expectation involved in the E-step.

## 1 Introduction

To date, various parametric families of densities have been proposed in the literature to represent the multivariate skew $t$ (MST) model, all having either a restricted or unrestricted characterization according to the classification scheme presented in Lee and McLachlan (2013b). In particular, the restricted MST (rMST) mixture model (or its equivalent variants by reparameterization) have been examined in Pyne et al. (2009), Cabral et al. (2012) and Vrbik and McNicholas (2012), while Lin (2010) and Lee and McLachlan (2011) focused on the unrestricted MST (uMST) mixture model adopting the characterization proposed by Sahu et al. (2003). In addition, Murray et al. (2013) recently considered an alternative construction that also adopted the term 'skew $t$-distribution', although referring to a quite different formulation more commonly known as the generalized-hyperbolic skew $t$ (GHST) mixture model. Although this latter distribution is a restricted type of skew distribution (under the argument that the skewing variable is a scalar), it differs from the rMST distribution in a number ways, such as having different behaviour in its tails and not becoming a skew normal distribution as a limiting and/or special case.

In this paper, we consider a natural and straightforward extension of the uMST distribution, where the skewing function of its density is given by a $q$-dimensional $t$-distribution function, and the skewness parameter is a general $p$ by $q$ matrix. This formulation coincides with a special case of the unified skew $t$ (SUT) distribution (Arellano-Valle and Azzalini, 2006) and the fundamental skew $t$ (FUST) distribution (Arellano-Valle and Genton, 2005), and is known as a location-scale variant of the canonical fundamental skew $t$ (CFUST) distribution. This characterization includes the restricted form of the MST distribution.

As discussed in Lee and McLachlan (2013a), while exact implementation of the Expectation-Maximization (EM) algorithm have been achievable for restricted characterizations of the component skew $t$-distributions, parameter estimation for the unrestricted and more general characterizations is substantially more intricate. In previous work on mixtures of unrestricted skew $t$-distributions using Sahu et al. (2003)'s characterization (a submodel of the CFUST distribution considered in this paper), Lin (2010) resorted to Monte Carlo methods for the computation of intractable conditional expectations involved in the E-step, while Lee and McLachlan (2013a) adopted the OSL (one-step late) approach to evaluating one of these difficult conditional expectation. In contrast for the rMST distribution, a number of closed-form implementations of the EM algorithm have been put forward (see Lee and McLachlan (2013a) for further discussion).

This paper presents an EM fitting algorithm for the FM-CFUST model. Under this general formulation, the aforementioned restricted and unrestricted skew $t$-mixture models can be obtained as special cases and, as such, the expressions on the E- and M-steps of the proposed algorithm would apply directly in these situations. In contrast to the approach of Lin (2010) which relies on computationally intensive numerical approximations, our development was motivated by the desire for an exact implementation of the EM algorithm using the truncated moments approach (Lee and McLachlan, 2013a, Ho et al., 2012), leading to much faster and more accurate results. Also, we derive a new exact expression for the remaining E-step conditional expectation that had previously relied on approximation or numerical methods.

## 2 The CFUST distribution

The multivariate canonical fundamental skew $t$-distribution (CFUST) was introduced by Arellano-Valle and Genton (2005). A location-scale variant of the CFUST distribution can be characterized as follows. Let $\boldsymbol{U}_1$ be $p$-dimensional random vector and $\boldsymbol{U}_0$ a $q$-dimensional random vector. Suppose that, conditional on a scalar gamma variable $w \sim \text{gamma}(\frac{\nu}{2}, \frac{\nu}{2})$, the joint distribution of $\boldsymbol{U}_0$ and $\boldsymbol{U}_1$ is given by

$$\begin{bmatrix} \boldsymbol{U}_0 \\ \boldsymbol{U}_1 \end{bmatrix} \sim N_{q+p}\left(\begin{bmatrix} \boldsymbol{0} \\ \boldsymbol{0} \end{bmatrix}, \frac{1}{w}\begin{bmatrix} \boldsymbol{I}_q & \boldsymbol{0} \\ \boldsymbol{0} & \boldsymbol{\Sigma} \end{bmatrix}\right), \qquad (1)$$

where $\nu$ is a scalar, $\boldsymbol{I}_q$ denotes the $q \times q$ identity matrix, $\boldsymbol{\Sigma}$ is a positive definite scale matrix, and $\boldsymbol{0}$ is a vector/matrix of zeros with appropriate dimensions. Then

$$\boldsymbol{Y} = \boldsymbol{\mu} + \boldsymbol{\Delta}\left|\boldsymbol{U}_0\right| + \boldsymbol{U}_1 \qquad (2)$$

follows the CFUST distribution, whose density is defined by

$$f\left(\boldsymbol{y}; \boldsymbol{\mu}, \boldsymbol{\Sigma}, \boldsymbol{\Delta}, \nu\right) = 2^q\, t_p\left(\boldsymbol{y}; \boldsymbol{\mu}, \boldsymbol{\Omega}, \nu\right) T_q\left(\boldsymbol{q}(\boldsymbol{y})\sqrt{\frac{\nu+p}{\nu+d(\boldsymbol{y})}}; \boldsymbol{0}, \boldsymbol{\Lambda}, \nu+p\right), \qquad (3)$$

where

$$\begin{aligned}
\boldsymbol{\Omega} &= \boldsymbol{\Sigma} + \boldsymbol{\Delta}\boldsymbol{\Delta}^T, \\
\boldsymbol{q} &= \boldsymbol{\Delta}^T\boldsymbol{\Omega}^{-1}\left(\boldsymbol{y} - \boldsymbol{\mu}\right), \\
\boldsymbol{\Lambda} &= \boldsymbol{I}_p - \boldsymbol{\Delta}^T\boldsymbol{\Omega}^{-1}\boldsymbol{\Delta}, \\
d(\boldsymbol{y}) &= (\boldsymbol{y} - \boldsymbol{\mu})^T \boldsymbol{\Omega}^{-1}(\boldsymbol{y} - \boldsymbol{\mu}).
\end{aligned}$$

Here, we let $t_p(\boldsymbol{y}; \boldsymbol{\mu}, \boldsymbol{\Omega}, \nu)$ denotes the $p$-dimensional $t$-distribution with location parameter $\boldsymbol{\mu}$, scale matrix $\boldsymbol{\Omega}$, and degree of freedom $v + p$, and $T_q(.)$ is the $q$-dimensional cumulative $t$-distribution function. In the above, $|\boldsymbol{U}_0|$ denotes the vector whose $i$th element is the magnitude of the $i$th element of the vector $\boldsymbol{U}_0$.

## 2.1 The unrestricted characterization as a special case

As mentioned previously, the characterization of Sahu et al. (2003) corresponds to the CFUST distribution with $q = p$ and $\boldsymbol{\Delta} = \text{diag}(\boldsymbol{\delta})$. Hence, we can write $\boldsymbol{\Lambda} = \boldsymbol{I}_p - \boldsymbol{\Delta}\boldsymbol{\Sigma}^{-1}\boldsymbol{\Delta}$, $\boldsymbol{q} = \boldsymbol{\Delta}\boldsymbol{\Omega}^{-1}(\boldsymbol{y} - \boldsymbol{\mu})$ and $\boldsymbol{\Omega} = \boldsymbol{\Sigma} + \boldsymbol{\Delta}^2$, and the uMST density is given by

$$f(\boldsymbol{y}; \boldsymbol{\mu}, \boldsymbol{\Sigma}, \boldsymbol{\delta}, \nu) = 2^p \, t_p(\boldsymbol{y}; \boldsymbol{\mu}, \boldsymbol{\Omega}, \nu) \, T_p\left(\boldsymbol{q}(\boldsymbol{y})\sqrt{\frac{\nu + p}{\nu + d(\boldsymbol{y})}}; \boldsymbol{0}, \boldsymbol{\Lambda}, \nu + p\right). \tag{4}$$

The uMST distribution has the same stochastic representation as given by (2) and (1) with $q$ replaced by $p$.

## 2.2 The restricted characterization as a special case

In the restricted characterization of the MST distribution, the convolution-type stochastic representation is given by

$$\boldsymbol{Y} = \boldsymbol{\mu} + \boldsymbol{\delta}|U_0| + \boldsymbol{U}_1 \tag{5}$$

where

$$\begin{bmatrix} U_0 \\ \boldsymbol{U}_1 \end{bmatrix} \sim N_{1+p}\left(\begin{bmatrix} 0 \\ \boldsymbol{0} \end{bmatrix}, \frac{1}{w}\begin{bmatrix} 1 & \boldsymbol{0} \\ \boldsymbol{0} & \boldsymbol{\Sigma} \end{bmatrix}\right). \tag{6}$$

It follows that the density of the rMST distribution is given by

$$f(\boldsymbol{y}) = 2t_p(\boldsymbol{y}; \boldsymbol{\mu}, \boldsymbol{\Omega}, \nu) T_1\left(\boldsymbol{\delta}^T \boldsymbol{\Omega}^{-1}(\boldsymbol{y} - \boldsymbol{\mu}), \left(\frac{\nu + d(\boldsymbol{y})}{\nu + p}\right)(1 - \boldsymbol{\delta}^T \boldsymbol{\Omega}^{-1} \boldsymbol{\delta}), \nu + p\right). \tag{7}$$

On directly comparing (5) and (2), it can be observed that the skewing effect of $\boldsymbol{\delta}|U_0|$ in the rMST distribution can be achieved by the CFUST formulation in a number of ways, including

- taking $\boldsymbol{\Delta} = \text{diag}(\boldsymbol{\delta})$ and in the unrestricted case, and setting $\boldsymbol{U}_0 = U_0 \boldsymbol{1}_p$ in (2), where $\boldsymbol{1}_p$ denotes a $p$-dimensional vector with elements one;

- constraining the skewness matrix $\boldsymbol{\Delta}$ to be a $p \times p$ zero matrix except for one column which is given by $\boldsymbol{\delta}$, and setting the corresponding element in $\boldsymbol{U}_0$ to be $U_0$; or

- setting $q = 1$.

As a corollary, it can be seen that the special case of $q = p$ of the CFUST distribution encompass both the rMST and uMST distribution through structural constraints on the skewness parameter. This is given by taking

$$\boldsymbol{\Delta} = \begin{bmatrix} \delta_1 & 0 & \ldots & 0 \\ \delta_2 & 0 & \ldots & 0 \\ \vdots & \vdots & \ddots & \vdots \\ \delta_p & 0 & \ldots & 0 \end{bmatrix} \text{ and } \boldsymbol{\Delta} = \begin{bmatrix} \delta_1 & 0 & \ldots & 0 \\ 0 & \delta_2 & \ldots & 0 \\ \vdots & \vdots & \ddots & \vdots \\ 0 & 0 & \ldots & \delta_p \end{bmatrix} \tag{8}$$

for the rMST and uMST distributions respectively, where $\boldsymbol{\delta} = (\delta_1, \delta_2, \ldots, \delta_p)^T$. Note that for the rMST case, $\boldsymbol{\delta}$ can be placed in any one of the columns of $\boldsymbol{\Delta}$, not necessarily the first column as given in (8).

# 3 Parameter estimation via the EM algorithm

We are interested in the finite mixtures with component densities given by (3). A $g$-component mixture of CFUST distributions has density

$$f(\boldsymbol{y}; \boldsymbol{\Psi}) = \sum_{h=1}^{g} \pi_h f(\boldsymbol{y}; \boldsymbol{\mu}_h, \boldsymbol{\Sigma}_h, \boldsymbol{\Delta}_h, \nu_h), \tag{9}$$

where $\pi_h$ $(h = 1, \ldots, g)$ are the mixing proportions and $f(.)$ denotes a CFUST density given by (3). We shall adopt the acronym FM-CFUST for (9). The vector $\boldsymbol{\Psi} = (\pi_1, \ldots, \pi_{g-1}, \boldsymbol{\theta}_1^T, \ldots, \boldsymbol{\theta}_g^T)$ contains all the unknown parameters of the model, with $\boldsymbol{\theta}_h$ containing the elements of $\boldsymbol{\mu}_h$ and $\boldsymbol{\delta}_h$, the distinct elements of $\boldsymbol{\Sigma}_h$, and $\nu_h$.

Similar to the rMST and uMST distributions, the CFUST admits a convenient hierarchical representation that greatly facilitates parameter estimation via the EM algorithm:

$$\begin{aligned}
\boldsymbol{Y} \mid \boldsymbol{U}, w &\sim N_p\left(\boldsymbol{\mu} + \boldsymbol{\Delta}\boldsymbol{u}, \frac{1}{w}\boldsymbol{\Sigma}\right) \\
\boldsymbol{U} \mid w &\sim HN_q\left(\boldsymbol{0}, \frac{1}{w}\boldsymbol{I}_m\right) \\
w &\sim gamma\left(\frac{\nu}{2}, \frac{\nu}{2}\right)
\end{aligned} \tag{10}$$

where $N_p(\boldsymbol{\mu}, \boldsymbol{\Sigma})$ denotes the multivariate normal distribution and $HN_q(0, \boldsymbol{\Sigma})$ represents the $q$-dimensional half-normal distribution with mean $\boldsymbol{0}$ and scale matrix $\boldsymbol{\Sigma}$. The EM algorithm for FM-CFUST then proceeds in a similar way as for the FM-uMST model as described in Lee and McLachlan (2013a).

## 3.1 E-Step

In the E-step, the following conditional expectations are computed:

$$\begin{aligned}
z_{hj}^{(k)} &= E_{\Psi^{(k)}}\left[z_{hj} = 1 \mid \boldsymbol{y}_j\right] \\
w_{hj}^{(k)} &= E_{\Psi^{(k)}}\left[w_{hj} \mid \boldsymbol{y}_j, z_{hj} = 1\right], \\
e_{1hj}^{(k)} &= E_{\Psi^{(k)}}\left[\log(w_{hj}) \mid \boldsymbol{y}_j, z_{hj} = 1\right], \\
\boldsymbol{e}_{2hj}^{(k)} &= E_{\Psi^{(k)}}\left[w_{hj}\boldsymbol{u}_{hj} \mid \boldsymbol{y}_j, z_{hj} = 1\right], \\
\boldsymbol{e}_{3hj}^{(k)} &= E_{\Psi^{(k)}}\left[w_{hj}\boldsymbol{u}_{hj}\boldsymbol{u}_{hj}^T \mid, \boldsymbol{y}_j, z_{hj} = 1\right].
\end{aligned}$$

They are given by

$$
\begin{aligned}
z_{hj}^{(k)} &= \frac{\pi_h f(\boldsymbol{y}_j; \boldsymbol{\mu}_h^{(k)}, \boldsymbol{\Sigma}_h^{(k)}, \boldsymbol{\delta}_h^{(k)}, \nu_h^{(k)})}{f(\boldsymbol{y}_j; \boldsymbol{\Psi}^{(k)})}, \\
w_{hj}^{(k)} &= \left(\frac{\nu_h^{(k)} + p}{\nu_h^{(k)} + d_h^{(k)}(\boldsymbol{y}_j)}\right) \frac{T_q\left(\boldsymbol{q}_h^{(k)} \sqrt{\frac{\nu_h^{(k)}+p+2}{\nu_h^{(k)} d_h^{(k)}(\boldsymbol{y}_j)}}; \boldsymbol{0}, \boldsymbol{\Lambda}_h^{(k)}, \nu_h^{(k)} + p + 2\right)}{T_q\left(\boldsymbol{q}_h^{(k)} \sqrt{\frac{\nu_h^{(k)}+p}{\nu_h^{(k)} d_h^{(k)}(\boldsymbol{y}_j)}}; \boldsymbol{0}, \boldsymbol{\Lambda}_h^{(k)}, \nu_h^{(k)} + p\right)}, \\
e_{1hj}^{(k)} &= w_{hj}^{(k)} - \log\left(\frac{\nu_h^{(k)} + d_h^{(k)}(\boldsymbol{y}_j)}{2}\right) + \left(\frac{\nu_h^{(k)} + p}{\nu_h^{(k)} + d_h^{(k)}(\boldsymbol{y}_j)}\right) - \psi\left(\frac{\nu_h^{(k)} + p}{2}\right) \quad (11) \\
\boldsymbol{e}_{2,hj}^{(k)} &= w_{hj}^{(k)} E_{\boldsymbol{\Psi}^{(k)}}[\boldsymbol{u}_{hj} \mid \boldsymbol{y}_j], \\
\boldsymbol{e}_{3hj}^{(k)} &= w_{hj}^{(k)} E_{\boldsymbol{\Psi}^{(k)}}[\boldsymbol{u}_{hj} \boldsymbol{u}_{hj}^T \mid \boldsymbol{y}_j], \quad (12)
\end{aligned}
$$

where $\boldsymbol{U}_{hj} \mid \boldsymbol{y}_j$ has a $q$-dimensional truncated $t$-distribution given by

$$
\boldsymbol{U}_{hj} \mid \boldsymbol{y}_j \sim tt_q\left(\boldsymbol{q}_h^{(k)}, \left(\frac{\nu_h^{(k)} + d_h(\boldsymbol{y}_j)}{\nu_h^{(k)} + p + 2}\right) \boldsymbol{\Lambda}_h^{(k)}, \nu_h^{(k)} + p + 2; \mathbb{R}^+\right). \quad (13)
$$

Note that (11) is obtained using a OSL EM algorithm. The derivations for these results are analogous to Lee and McLachlan (2013a). Alternatively, a series truncation approach can be used in place of (11), which exploits an exact representation of this conditional expectation, will be presented later.

## 3.2 M-Step

On the $(k+1)$th M-step, the model parameters are updated with

$$
\begin{aligned}
\pi_h^{(k+1)} &= \frac{1}{n} \sum_{j=1}^n z_{hj}^{(k)}, \\
\boldsymbol{\mu}_h^{(k+1)} &= \frac{\sum_{j=1}^n z_{hj} w_{hj}^{(k)} \boldsymbol{y}_j - \boldsymbol{\Delta}_h^{(k)} \sum_{j=1}^n z_{hj}^{(k)} \boldsymbol{e}_{2hj}^{(k)}}{\sum_{j=1^n} z_{hj}^{(k)} w_{hj}^{(k)}}, \\
\boldsymbol{\Delta}_h^{(k+1)} &= \left[\sum_{j=1}^n z_{hj}^{(k)}\left(\boldsymbol{y}_j - \boldsymbol{\mu}_h^{(k+1)}\right) \boldsymbol{e}_{2hj}^{(k)T}\right] \left[\sum_{j=1}^n z_{hj}^{(k)} \boldsymbol{e}_{3hj}^{(k)}\right]^{-1}, \quad (14) \\
\boldsymbol{\Sigma}_h^{(k+1)} &= \left[\sum_{j=1}^n z_{hj}^{(k)}\right]^{-1} \Bigg\{\sum_{j=1}^n z_{hj}^{(k)}\left[w_{hj}^{(k)}\left(\boldsymbol{y}_j - \boldsymbol{\mu}_h^{(k+1)}\right)\left(\boldsymbol{y}_j - \boldsymbol{\mu}_h^{(k+1)}\right)^T - \boldsymbol{\Delta}_h^{(k+1)} \boldsymbol{e}_{2hj}^{(k)}\left(\boldsymbol{y}_j - \boldsymbol{\mu}_h^{(k+1)}\right)^T\right. \\
&\quad \left.- \left(\boldsymbol{y}_j - \boldsymbol{\mu}_h^{(k+1)}\right) \boldsymbol{e}_{2hj}^{(k)T} \boldsymbol{\Delta}_h^{(k+1)T} + \boldsymbol{\Delta}_h^{(k+1)} \boldsymbol{e}_{3hj}^{(k)T} \boldsymbol{\Delta}_h^{(k+1)T}\right]\Bigg\}. \quad (15)
\end{aligned}
$$

An update of the degrees of freedom is obtained by solving for $\nu_h$:

$$
0 = \left(\sum_{h=1}^n z_{hj}^{(k)}\right)\left[\log\left(\frac{\nu_h}{2}\right) - \psi\left(\frac{\nu_h}{2}\right) + 1\right] - \left(\sum_{j=1}^n e_{1hj}^{(k)} - w_{hj}^{(k)}\right).
$$

## 3.3 The rMST and uMST as special cases

As pointed out previously, the rMST and uMST distributions can be obtained as special cases of the CFUST distribution by imposing structural constraints on the skewness parameter $\boldsymbol{\Delta}$. In the case of the uMST distribution, the diagonal elements of the component skewness matrices are estimated as

$$\boldsymbol{\delta}_h^{(k+1)} = \left(\boldsymbol{\Sigma}_h^{(k)-1} \circ \sum_{j=1}^n \tau_{hj}^{(k)} \boldsymbol{e}_{4,hj}^{(k)}\right)^{-1} \text{diag}\left[\boldsymbol{\Sigma}_h^{(k)-1} \sum_{j=1}^n \tau_{hj}^{(k)} \left(\boldsymbol{y}_j - \boldsymbol{\mu}_h^{(k+1)}\right) \boldsymbol{e}_{3,hj}^{(k)T}\right], \quad (16)$$

where $\circ$ denotes the elementwise product.

Under the constraint that $\boldsymbol{\Delta}_h$ is zero except for the first column $\boldsymbol{\delta}_h$, that is, $\boldsymbol{\Delta} = [\boldsymbol{\delta}\, \boldsymbol{0} \ldots \boldsymbol{0}]$, (14) reduces to estimating

$$\boldsymbol{\delta}_h^{(k+1)} = \frac{\sum_{j=1}^n \tau_{hj}^{(k)} [\boldsymbol{e}_{3,hj}^{(k)}]_1 (\boldsymbol{y}_j - \boldsymbol{\mu}_h^{(k+1)})}{\sum_{j=1}^n \tau_{hj}^{(k)} [\boldsymbol{e}_{4,hj}^{(k)}]_{11}}, \quad (17)$$

where $[\boldsymbol{e}_{3,hj}^{(k)}]_1$ and $[\boldsymbol{e}_{4,hj}^{(k)}]_{11}$ denote the first element of $\boldsymbol{e}_{3,hj}^{(k)}$ and $\boldsymbol{e}_{4,jh}^{(k)}$, respectively. On comparing the above with the corresponding expression for the rMST mixture model (see, for example, equation (9) from Wang et al. (2009)), it can observed that the two expressions are indeed equivalent.

## 3.4 New results on $e_{1,hj}^{(k)}$: the truncated series approach

We now present the derivation of a new exact expression for $e_{1,hj}^{(k)}$, expressed in terms of an infinite series involving multivariate $t$-distribution functions. By the law of iterated expectations,

$$\begin{aligned}
e_{1,hj}^{(k)} &= E_{\boldsymbol{\Psi}^{(k)}}\left[\log(W_j) \mid \boldsymbol{y}_j\right] \\
&= E_{\boldsymbol{\Psi}^{(k)}}\left[E_{\boldsymbol{\Psi}^{(k)}}\left[\log(W_j) \mid \boldsymbol{u}_j, \boldsymbol{y}_j\right] \mid \boldsymbol{y}_j\right] \\
&= E_{\boldsymbol{\Psi}^{(k)}}\left[\psi\left(\frac{\nu_h^{(k)}+p}{2}\right) - \log\left(\frac{\nu_h^{(k)} + d_h^{(k)}(\boldsymbol{y}_j) + (\boldsymbol{u}_j - \boldsymbol{q}_{hj}^{(k)})^T \boldsymbol{\Lambda}_h^{(k)-1}(\boldsymbol{u}_j - \boldsymbol{q}_{hj}^{(k)})}{2}\right) \mid \boldsymbol{y}_j\right] \\
&= \psi\left(\frac{\nu_h^{(k)}+p}{2}\right) - \log\left(\frac{\nu_h^{(k)} + d_h^{(k)}(\boldsymbol{y}_j)}{2}\right) - E_{\boldsymbol{\Psi}^{(k)}}\left[\log\left(1 + \frac{(\boldsymbol{u}_j - \boldsymbol{q}_{hj}^{(k)})^T \boldsymbol{\Lambda}_h^{(k)-1}(\boldsymbol{u}_j - \boldsymbol{q}_{hj}^{(k)})}{\nu_h^{(k)} + d_h^{(k)}(\boldsymbol{y}_j)}\right)\right].
\end{aligned}$$
(18)

To calculate the last term in (18), note that the conditional density of $\boldsymbol{u}_j$ given $\boldsymbol{y}_j$ can be written as

$$f(\boldsymbol{u}_j \mid \boldsymbol{y}_j) = \frac{t_q\left(\boldsymbol{u}_j; \boldsymbol{q}_{hj}^{(k)}, \left(\frac{\nu_h^{(k)} + d_h^{(k)}(\boldsymbol{y}_j)}{\nu_h^{(k)}+p}\right)\boldsymbol{\Lambda}_h^{(k)}, \nu_h^{(k)}+p\right)}{T_q\left(\boldsymbol{q}_{hj}^{(k)}; \boldsymbol{0}, \left(\frac{\nu_h^{(k)} + d_h^{(k)}(\boldsymbol{y}_j)}{\nu_h^{(k)}+p}\right)\boldsymbol{\Lambda}_h^{(k)}, \nu_h^{(k)}+p\right)}. \quad (19)$$

After some algebraic manipulations, it follows that the expectation in (18) can be reduced to a simple closed-form expression (see equation (76) from Lee and McLachlan (2013a)), except for a term involving an integral given by

$$
\begin{aligned}
S_{1,hj}^{(k)} &= \int_0^\infty \log\left(1 + \frac{(\boldsymbol{u}_j - \boldsymbol{q}_{hj}^{(k)})^T \boldsymbol{\Lambda}_h^{(k)^{-1}}(\boldsymbol{u}_j - \boldsymbol{q}_{hj}^{(k)})}{\nu_h^{(k)} + d_h^{(k)}(\boldsymbol{y}_j)}\right) \\
&\quad t_q\left(\boldsymbol{u}_j; \boldsymbol{q}_{hj}^{(k)}, \left(\frac{\nu_h^{(k)} + d_h^{(k)}(\boldsymbol{y}_j)}{\nu_h^{(k)} + p}\right) \boldsymbol{\Lambda}_h^{(k)}, \nu_h^{(k)} + p\right) d\boldsymbol{u}_j \\
&= \sum_{r=1}^\infty \sum_{s=0}^r \frac{(-1)^{2r-s-1}}{r} \binom{r}{s} \int_0^\infty \left(1 + \frac{(\boldsymbol{u}_j - \boldsymbol{q}_{hj}^{(k)})^T \boldsymbol{\Lambda}_h^{(k)^{-1}}(\boldsymbol{u}_j - \boldsymbol{q}_{hj}^{(k)})}{\nu_h^{(k)} + d_h^{(k)}(\boldsymbol{y}_j)}\right)^{-s} \\
&\quad t_q\left(\boldsymbol{u}_j; \boldsymbol{q}_{hj}^{(k)}, \left(\frac{\nu_h^{(k)} + d_h^{(k)}(\boldsymbol{y}_j)}{\nu_h^{(k)} + p}\right) \boldsymbol{\Lambda}_h^{(k)}, \nu_h^{(k)} + p\right) d\boldsymbol{u}_j \qquad (20) \\
&= \sum_{r=1}^\infty \sum_{s=0}^r \frac{(-1)^{2r-s-1}}{r} \binom{r}{s} \frac{\Gamma\left(\frac{\nu_h^{(s)}+p}{2} + s\right)}{\Gamma\left(\frac{\nu_h^{(k)}+p}{2}\right)} \frac{\Gamma\left(\frac{\nu_h^{(k)}+p+q}{2}\right)}{\Gamma\left(\frac{\nu_h^{(k)}+p+q}{2} + s\right)} \\
&\quad \int_0^\infty t_q\left(\boldsymbol{u}_j; \boldsymbol{q}_{hj}^{(k)}, \left(\frac{\nu_h^{(k)} + d_h^{(k)}(\boldsymbol{y}_j)}{\nu_h^{(k)} + p + 2s}\right) \boldsymbol{\Lambda}_h^{(k)}, \nu_h^{(k)} + p + 2s\right) d\boldsymbol{u}_j \qquad (21) \\
&= \frac{\Gamma\left(\frac{\nu_h^{(k)}+p+q}{2}\right)}{\Gamma\left(\frac{\nu_h^{(k)}+p}{2}\right)} \sum_{r=1}^\infty \sum_{s=0}^r \frac{(-1)^{2r-s-1}}{r} \binom{r}{s} \frac{\Gamma\left(\frac{\nu_h^{(s)}+p}{2} + s\right)}{\Gamma\left(\frac{\nu_h^{(k)}+p+q}{2} + s\right)} \\
&\quad T_q\left(\boldsymbol{q}_{hj}^{(k)}; \boldsymbol{0}, \left(\frac{\nu_h^{(k)} + d_h^{(k)}(\boldsymbol{y}_j)}{\nu_h^{(k)} + p + 2s}\right) \boldsymbol{\Lambda}_h^{(k)}, \nu_h^{(k)} + p + 2s\right) \qquad (22)
\end{aligned}
$$

In (20) above, we have applied a Taylor expansion of $\log(x)$ for $0 < x < 2$ and the binomial expansion formula to obtain a series expression for $\log(x)$, given by

$$\log(x) = \sum_{r=1}^\infty \sum_{s=0}^r \frac{(-1)^{2r-s-1}}{r} \binom{r}{s} x^s, \quad \text{for } 0 < x < 2.$$

Note that the term inside the logarithm in (21) is always greater than one. Hence, its inverse lies between 0 and 1, and we can apply the above series representation. In practice, $S_{1,hj}^{(k)}$ can be approximated by a (small) finite number of terms.

It follows that $e_{1,hj}^{(k)}$ is given by

$$
\begin{aligned}
e_{1,hj}^{(k)} &= \psi\left(\frac{\nu_h^{(k)}+p}{2}\right) - \log\left(\frac{\nu_h^{(k)}+d_h^{(k)}(\boldsymbol{y}_j)}{2}\right) \\
&+ \frac{\Gamma\left(\frac{\nu_h^{(k)}+p+q}{2}\right)}{\Gamma\left(\frac{\nu_h^{(k)}+p}{2}\right)} T_q^{-1}\left(\boldsymbol{q}_{hj}^{(k)}; \boldsymbol{0}, \left(\frac{\nu_h^{(k)}+d_h^{(k)}(\boldsymbol{y}_j)}{\nu_h^{(k)}+p}\right)\boldsymbol{\Lambda}_h^{(k)}, \nu_h^{(k)}+p\right) \\
&\sum_{r=1}^{\infty}\sum_{s=0}^{r} \frac{(-1)^{2r-s-1}}{r} \binom{r}{s} \frac{\Gamma\left(\frac{\nu_h^{(s)}+p}{2}+s\right)}{\Gamma\left(\frac{\nu_h^{(k)}+p+q}{2}+s\right)} \\
&T_q\left(\boldsymbol{q}_{hj}^{(k)}; \boldsymbol{0}, \left(\frac{\nu_h^{(k)}+d_h^{(k)}(\boldsymbol{y}_j)}{\nu_h^{(k)}+p+2s}\right)\boldsymbol{\Lambda}_h^{(k)}, \nu_h^{(k)}+p+2s\right).
\end{aligned} \quad (23)
$$

# 4 Conclusions

We have introduced a finite mixture of CFUST distributions, where the component densities are the generalization of the restricted and unrestricted skew $t$-distributions. Parameter estimation via the EM algorithm has been outlined, following the truncated moments approach in Ho et al. (2012) and Lee and McLachlan (2013a). Furthermore, new results have been derived for conditional expectation $w_{hj}^{(k)}$, exploiting an infinite series representation of the logarithm function. This leads to a full implementation of the EM algorithm for the FM-CFUST model with exact and closed-form expressions for all E-step conditional expectations. Simulations and applications of the FM-CFUST model, as well as extension to factor analyzers, will be treated in a forthcoming manuscript.